\documentstyle[12pt,epsfig]{article} 

\begin{document}

\begin{center}
{\huge Strangeness Production in Relativistic Heavy Ion Experiments\\}
\vspace*{1.5cm}
{\large Stathes D. Paganis
\footnote{e-mail address: paganis@physics.utexas.edu}\\
         Department of Physics, University of Texas at Austin\\
         Austin, Texas 78712}
\end{center}
\vspace*{2cm}

\begin{abstract}
The role of Strangeness as a signal of the Quark Gluon Plasma in 
relativistic heavy ion experiments is discussed. 
The current experimental status is briefly presented. Several scenarios 
which explain the CERN data are discussed.
\end{abstract}
\section{Introduction}
\vspace{0.2in}
QCD, the quantum field theory of the strong interactions,
is expected to contain two very interesting non-perturbative 
effects in its low energy regime: Confinement and Chiral 
Symmetry Breaking (CSB). Since the QCD degrees of freedom 
are quarks and gluons which in the present universe temperature 
are confined in the observed hadrons, 
it is widely accepted that strongly interacting matter 
at low energy density is sufficiently described by massive 
hadrons. These are the effective degrees of freedom of many 
low energy effective field theories of the strong interactions. 
From the hadron spectrum and the very low masses of the 
pseudoscalar mesons (pions and kaons) it is expected that QCD 
in its massless quark limit exhibits spontaneous symmetry breaking
of its Chiral Symmetry at low temperatures. 
The pseudoscalar mesons are the Goldstone bosons of this 
spontaneous symmetry breaking and the hadronic masses (like 
the proton mass) mainly come from the quark condensate 
$\langle \overline{q} q \rangle$ which gets a non 
zero value below the critical temperature.

It is expected that nuclear matter under high enough energy 
density undergoes a transition to a new form of matter 
where the degrees of freedom are free, almost massless 
quarks and gluons. This new form of matter has been proposed 
long ago \cite{collins} and is called the Quark Gluon Plasma (QGP).
This expectation is motivated by a series of lattice QCD results 
\cite{lattice} 
and has not been theoretically proven since QCD has not been 
solved. Consequently experiment is needed to settle the 
question. Relativistic heavy ion collisions are used as 
an experimental tool to study strongly interacting matter 
under high energy and baryon densities. In such experiments 
one hopes that if a QGP is formed, then certain 
experimental observables will be measured which might be 
significantly different than the observables measured when 
a Hadronic Gas (HG) is formed.
These observables are usually called signals or signatures of 
the QGP.

In this talk I will focus on a dinstinctive signal of QGP, the 
predicted enhancement of strange particle production. 
In the first section the use of strangeness production as 
a QGP signal is justified. In the second section 
experimental results are presented and various theoretical 
explanations are briefly discussed.
The transition from 
a HG to QGP actually involves two transitions, one related to 
confinement-deconfinement and the other to 
chiral symmetry restoration. 
Typically in the literature 
the critical temperature for both transitions is predicted to be 
the same \cite{lattice},\cite{harris}. 
In the last section of this paper we will sketch what 
could happen in a different scenario where the two transitions 
occur at significantly different critical temperatures.

\section{Why Strangeness?}
\vspace{0.2in}
Strange particle production in relativistic heavy ion collisions 
is interesting because it may provide a signal of QGP formation
and because of the possible existence of new exotic forms of matter.
In this section the strangeness enhancement as a signal of QGP 
formation is discussed. The hypothesized existence of 
a stable strange matter is briefly mentioned.

During a relativistic heavy ion collision a space-time region 
containing many particles with small mean free path is
formed, the so-called fireball \cite{eggers}. The fireball can 
be described in terms of thermodynamic variables such as 
temperature. As it was mentioned in the introduction at 
a high enough 
temperature a transition from a HG to QGP is expected.
A higher strangeness production (in terms of
$s \overline s$ pairs) is expected if a QGP forms with respect to 
a HG, usually referred to as strangeness enhancement. I 
discuss three main factors responsible for the stangeness 
enhancement: the kinematical factor, medium effects and 
the short fireball lifetime.

The principal reason for enhanced strangeness production in a 
QGP relative to that in a HG is due to the respective 
kinematic thresholds. 
In a hadron gas the threshold for the following typical reaction,
\begin{eqnarray}
p+p \rightarrow \Lambda + K + p
\end{eqnarray}
is about $700 MeV$, while in a QGP strangeness production occurs via 
the following reactions,
\begin{eqnarray}
q\overline q \rightarrow s\overline s \\
gg \rightarrow s\overline s 
\end{eqnarray}
with a corresponding threshold of $2m_s \simeq 300 MeV$ since the bare 
mass of the strange quark is about $150 MeV$. The gluon channel is 
the dominant one: while gluons are confined within 
hadrons in a HG, in QGP they are free within the fireball leading 
to a large cross section for this process \cite{koch}.

Medium effects are also responsible for strangeness enhancement. 
In QGP the ratio of $\overline s$ 
quarks with respect to $\overline u + \overline d$ is expected 
to be much higher than in a HG. The reason is that in a HG there 
is always some finite initial baryon density (because of the 
protons and neutrons that are stopped during the collision) which 
suppresses further production of $N \overline N$ 
pairs (Pauli blocking is in effect and more energy is required 
to create a nucleon). In this case a low 
$\overline \Lambda$ phase space density is expected 
because the $\overline \Lambda$ production proceeds 
via the abundance of 
antinucleons which are already suppressed \cite{koch2}. 
On the other hand in a QGP the $s \overline s$ 
production is only suppressed due to the strange quark mass 
(there is no net initial strangeness in the fireball). So,
an antistrangeness enhancement is expected in QGP. Experimentally 
we expect the $\overline \Lambda / \overline N$ ratio to increase.

The lifetime of the fireball is very short (c$\tau \simeq 3-5 fm$)
with respect to the weak interaction time scale, so that 
strangeness can be considered as a conserved quantum number. Because 
of this short lifetime, strangeness equilibration is 
very questionable. By strangeness equilibrium we mean that the 
fireball system (either in a HG or in a QGP phase) has reached 
a state in which the strange quark density has reached a saturation density 
$\rho_s$ after a time $\tau_s$ \cite{singh}. The degree of 
strangeness equilibration $\gamma_s$ during the lifetime of 
the fireball determines the amount of final 
strangeness production in a 
heavy ion collision. Theoretical calculations \cite{koch} 
have shown that strangeness chemical equilibration is 10-30 
times slower in a HG at the same temperature and baryon 
density than in QGP where $c\tau_s \simeq 3 fm$. This means 
that during the fireball lifetime of a QGP all the available 
(from phase space) strangeness is produced while for a HG much
less is produced. This results in expected strangeness
enhancement in the case of QGP formation and makes strangeness
production a good signal of the HG $\rightarrow$ QGP transition.

Finally understanding strangeness production and evolution in 
heavy ion experiments is also important for more exotic reasons;
the hypothesized  stability of strange matter by Witten \cite{witten}
and the possible existence of strangelets \cite{fahri}, 
or multiquark hadrons such as the $H$ dibaryon \cite{jaffe}.
These phenomena have cosmological implications such as the 
existence of strange stars, understanding of the nature of 
the dark matter and others.

\section{Current Experimental Status}
\vspace{0.2in}

Experimentally strangeness production is measured in the form of 
various strange to non-strange particle ratios; the most 
common being the $K/\pi$ ratio. The strangeness suppression 
factor $\lambda_s$ is defined as \cite{wrobl}:
\begin{equation}
\lambda_s = \frac{\langle s \overline s \rangle}
{\frac{1}{2}(\langle u \overline u \rangle + \langle d 
\overline d \rangle )}.
\end{equation}
This ratio can be approximated by actual particle ratios; 
in experiment NA35 at the CERN-SPS the following approximation
was chosen \cite{na35}:
\begin{equation}
\lambda_s \simeq \frac{\langle \Lambda \rangle + 
4 \langle K^0_s \rangle} {3 \langle \pi^- \rangle}.
\end{equation}
This ratio was measured to be around $0.35$ in Sulphur+Sulphur 
collisions and $0.15-0.2$ in Nucleon-Nucleon collisions. The 
conclusion is that there is a significant strangeness 
enhancement as one goes from N+N to S+S collisions.
An increase in the multistrange particle yields was also 
observed in both CERN and AGS-BNL \cite{heinz}.

\begin{figure}[lbar]
\psfig{figure=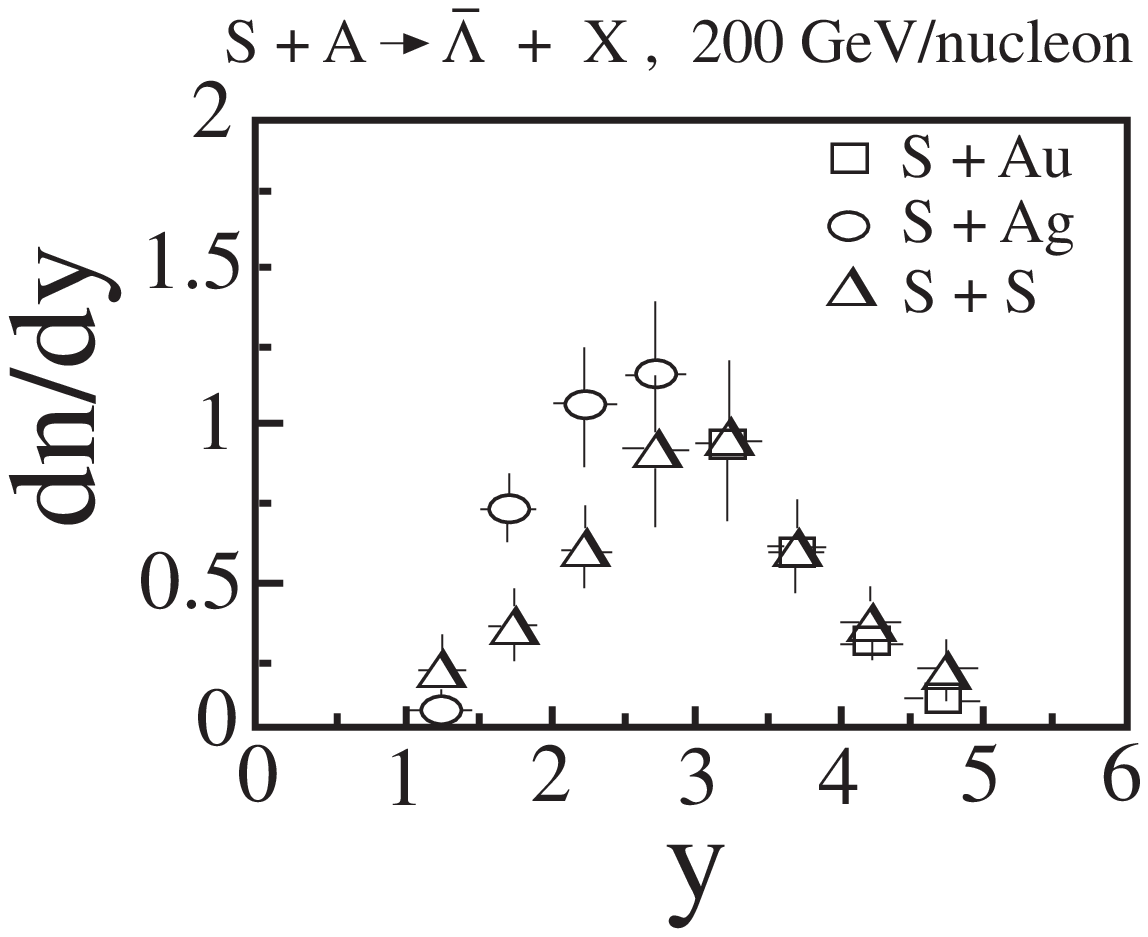,width=6.2in}
\caption{
Rapidity distribution of $\overline \Lambda$ produced in 
various central S+A collisions \protect\cite{spiros}.
}
\label{lbar}
\end{figure}

\begin{figure}[ratio1]
\psfig{figure=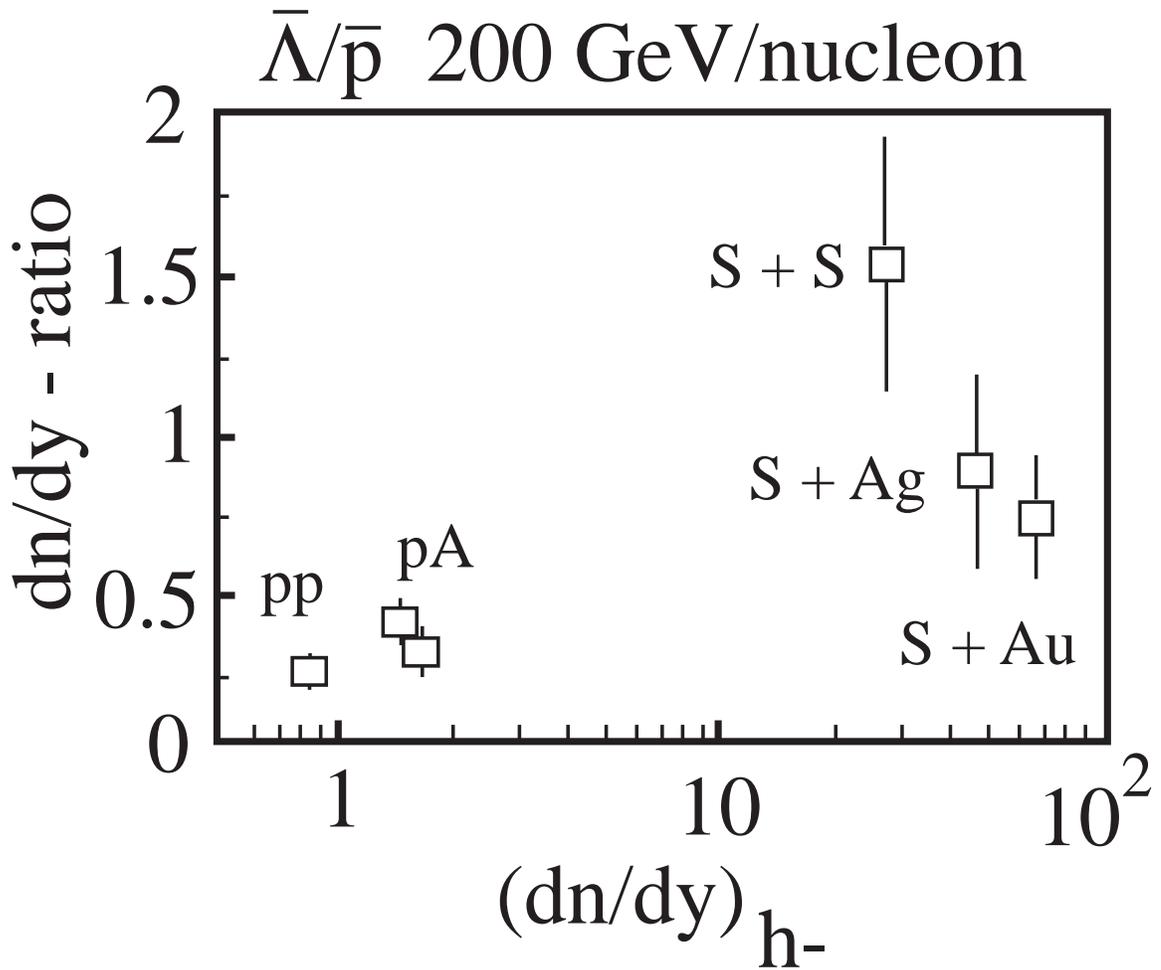,width=6.2in}
\caption{
$\overline \Lambda / \overline p$ ratio near midrapidity in
nucleon-nucleon, minimum bias proton-nucleus and central
nucleus-nucleus collisions as a function of the rapidity at
midrapidity of negatively charged hadrons \protect\cite{spiros}.
}
\label{ratio1}
\end{figure}

\begin{figure}[ratio2]
\psfig{figure=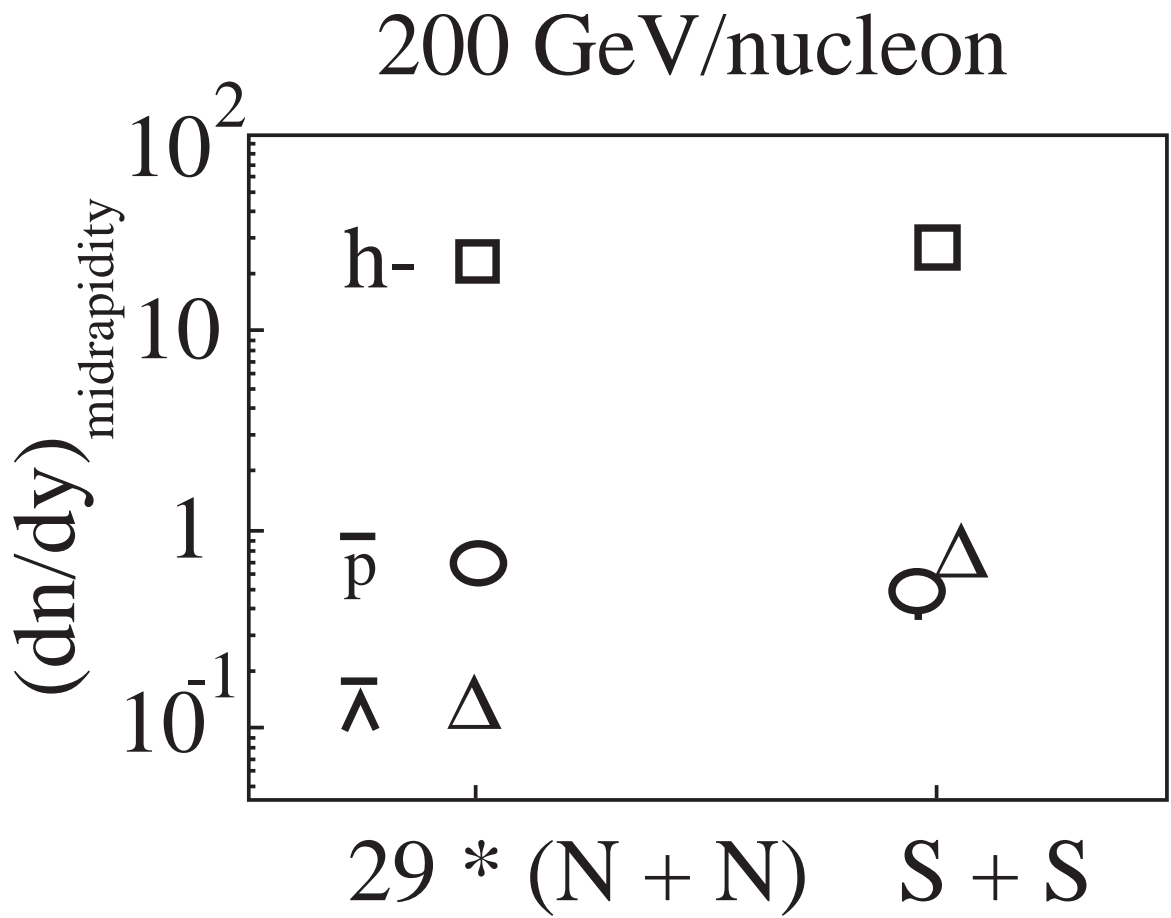,width=6.2in}
\caption{
$ \overline \Lambda$ and $ \overline p$ production near midrapidity in
central S+S collisions compared to nucleon-nucleon data scaled by the
corresponding pion multiplicity ratio in full 
phase space \protect\cite{spiros}.
}
\label{ratio2}
\end{figure}

Another very interesting ratio is the 
$\overline \Lambda (\overline u \overline d \overline s)/
\overline p (\overline u \overline u \overline d)$ ratio. 
As we said in the previous section if a QGP is formed this ratio
is expected to increase dramatically. In figures ~\ref{lbar},
~\ref{ratio1} and ~\ref{ratio2} the 
$\overline \Lambda$ production and the 
$\overline \Lambda / \overline p $ 
are shown for various collisions, ranging from $p+p$ to $S+Au$ 
\cite{spiros}.
A remarkable increase in the ratio is observed especially for 
the $S+S$ collisions. As fig.~\ref{ratio2} shows, this increase is mainly 
due to the $\overline \Lambda$ increase or equivalently due to an 
increase of the $\overline s$ production.

Thermal statistical models were used to analyze the CERN 
data \cite{heinz}. Their results suggest that there is no QGP at 
CERN and the large strangeness equilibration obtained, 
$\gamma_s \simeq 60-70 \% $ is only due to some strangeness
production mechanisms present in the hadronic level. These data 
can be explained by microscopic models that do not require a QGP.

A different but not as popular point of view 
suggests that there is 
a transition from a HG at AGS to a QGP at CERN \cite{gadz1}.
According to this view the experimental results indicate an 
increase of the effective degrees of freedom between BNL and 
CERN. The observables used in this study were entropy and 
strangeness.

\section{Future}
\vspace{0.2in}

The current understanding of the data from heavy 
ion experiments is that the QGP phase has not yet been 
formed and higher energies are needed; the current CERN 
SPS energies are $\sqrt{s} \simeq 20 AGeV$. In 1999 the 
new heavy ion collider RHIC at BNL New York will begin 
running and the STAR and PHENIX experiments will start data collection 
\cite{lanny}. The energy involved is $\sqrt{s} \simeq 200 AGeV$, 
i.e. one order of magnitude higher than CERN and is expected to 
put the system above the critical temperature.

Some of the advantages of higher RHIC energy with respect to 
measurement of strange particle production are:
\begin{itemize}
\item
Higher expected fireball temperature.
\item
Longer fireball lifetime is expected to allow higher 
strangeness chemical equilibration.
\end{itemize}

The main advantages of the STAR detector are:
\begin{itemize}
\item
Large acceptance. STAR covers a large portion of the phase space 
providing very high statistics even in an event by event basis.
As a result various important observables can be measured in an 
event by event basis.
\item
Good low $p_{\perp}$ coverage. The silicon vertex tracker SVT is 
capable of tracking pions with momenta as low as $50 MeV/c$.
\item
Good reconstruction of short lived singly strange baryons and 
mesons, and with the SVT, multiply strange baryons and antibaryons.
\end{itemize}
STAR will collect a number of observables for the same set of events
and will try to make a statement about the formation of 
QGP based on dramatic changes in these observables when some 
thermodynamic parameters (like temperature) change.

Finally, a future CERN heavy ion experiment, ALICE in LHC has been 
approved to run in year 2005. The energy is very high 
$\sqrt{s} \simeq 6 ATeV$ thus providing even better chances 
for producing QGP.

\section{Epilogue}
\vspace{0.2in}

In the quest for the experimental proof of the predicted QCD 
transition from 
HG to QGP, strangeness production is one of the most important 
observables. A significant strangeness enhancement has already been 
observed for high energy heavy ion collisions at CERN-SPS. 
This observation 
together with the very recent and exciting observation of 
$J/\psi$ suppression at CERN (\cite{na50a},\cite{na50b}) and the 
unexpectedly high pion (i.e. entropy) production at CERN
compared to the lower AGS, motivated speculation 
about possible plasma creation at SPS energies. This situation is 
reviewed in \cite{heinz} where it was stated that thermal 
statistical hadron gas models fit the CERN results for
large  strangeness equilibration $\gamma_s \simeq 60-70\%$. The 
large strangeness production and high pion 
multiplicities are attributed to hadronic mechanisms.
On the other hand the observed $J/\psi$ suppression should also 
be due to some hadronic mechanism if the HG scenario is the correct 
one for CERN-SPS energies \cite{gavin}.

The models described above assume that the two QCD transitions occur
at the same temperature. This assumption has no direct justification. 
In fact, some time ago Manohar and Georgi \cite{georgi}
suggested that the successes
of the non-relativistic quark model and the smallness of the 
effective strong coupling $\alpha_s$ are due to the two scales of QCD:
$\Lambda_{QCD} \simeq 200 MeV$ and $\Lambda_{\chi SB} \simeq 1 GeV$, 
for the confinement-deconfinement and chiral symmetry breaking scales
in zero temperature field theory.
The success of QCD sum rules also suggest this picture.
Based on this chiral quark picture there is a region between the 
two scales where the effective degrees of freedom are constituent 
quarks, gluons and Goldstone bosons. 
Phenomenologically this means that in this 
region although chiral symmetry is broken and the non-perturbative 
effects are significant, confinement has not set in yet. 
I believe that this opens 
a new possibility for the heavy ion experiments: the chiral symmetry 
might be harder to restore than to produce a deconfined phase. 
If this is the case in the CERN energies, then one should expect some 
moderate strangeness enhancement, a jump in the energy and entropy 
densities because of the change in the degrees of freedom and 
deconfinement effects like the observed $J/\psi$ suppression.
The author is currently investigating what the observable 
predictions of such a phase (deconfined but not chirally restored) are.

We close our discussion stressing that there are many physics pictures
that could explain the current experimental data. More experimental 
points in different energies (lower and higher) are needed. This 
additional data will be provided by future experiments at RHIC and CERN.

This work was supported in part by the U.S. Department of Energy and
the Robert A. Welch Foundation.

\end{document}